\documentclass[12pt]{article}
\usepackage{graphicx}
\begin{document}
%%----------------------------------------------------------------------------

\begin{center}

{\Large \bf Entropy and Temperature from  Entangled Space and Time}
\\[10mm]
Young S. Kim \\
Center for Fundamental Physics, University of Maryland,\\ College Park,
Maryland 20742, U.S.A. \\e-mail: yskim@umd.edu   \\[5mm]

    Marilyn E. Noz \\[1mm]
Department of Radiology, New York University \\ New York, New York, 10016,
U.S.A. \\e-mail: marilyne.noz@gmail.com

\end{center}

\vspace{20mm}

\abstract{Two coupled oscillators provide a mathematical instrument for
solving many problems in modern physics, including squeezed states of
light and Lorentz transformations of quantum bound states.  The concept
of entanglement can also be studied within this mathematical framework.
For the system of two entangled photons, it is of interest to study
what happens to the remaining photon if the other photon is not observed.
It is pointed out that this problem is an issue of Feynman's rest of
the universe.
For quantum bound-state problems, it is pointed out the longitudinal and
time-like coordinates become entangled when the system becomes boosted.
Since time-like oscillations are not observed, the problem is exactly
like the two-photon system where one of the photons is not observed.
While the hadron is a quantum bound state of quarks, it appears
quite differently when it moves rapidly than when it moves slowly.
For slow hadrons, Gell-Mann's quark model is applicable, while Feynman's
parton model is applicable to hadrons with their speeds close to that of
light.  While observing the temperature dependence of the speed, it is
possible to explain the quark-to-parton transition as a phase transition.}

\vspace{35mm}

\noindent  to be published in the Physical Science International Journal.

\newpage
\section{Introduction}\label{intro}

When Einstein developed his special relativity in 1905, he worked
out a transformation law for a point particle.  Presumably he was
aware that the hydrogen atom consists of one electron circling
around a proton.  However, he did not mention the question of
how the electron orbit appears to a moving observer.

\par
\begin{figure}[thb]
\centerline{\includegraphics[scale=1.5]{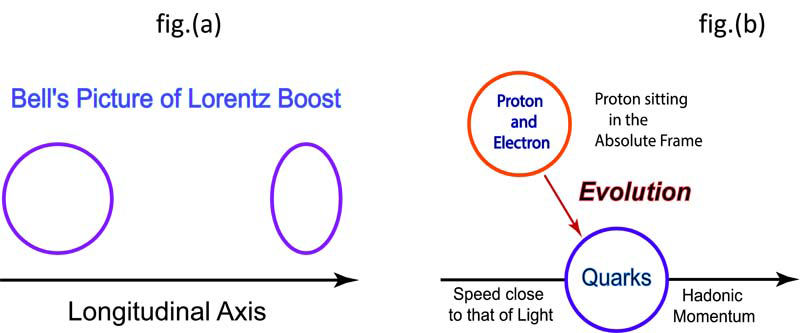}}
\vspace{5mm}
\caption{Hydrogen atom or standing wave in the Lorentz-covariant world.
The Lorentz contraction leads us to sketch an elliptic electron orbit
for the moving hydrogen atom~\cite{jsbell88}, as shown in fig.(a).
This became an outdated concept after 1927.  The electron orbit became
a standing wave.  The issue is how to Lorentz-boost the standing
wave.   Since the hydrogen atom cannot be accelerated, we have to
study the standing wave for the proton, as illustrated in fig.(b).
The proton is a bound state of quarks.  The energy of the proton
from CERN's LHC is 4,000 times its mass, and its speed is extremely
close to that of light.}\label{evol}
\end{figure}

Since then, many authors had ideas about a possible elliptic
deformation of the orbit as described in Fig.~\ref{evol}.  However,
quantum mechanics made this concept obsolete.  The hydrogen orbit
was replaced by a standing wave.  Thus, the question is how
the standing wave looks to a moving observer.

\par
For thirty six years from 1927 to 1963, Paul A. M. Dirac was occupied
with this problem.  In 1927~\cite{dir27}, Dirac noted that there is an
uncertainty relation between time and energy variables but there are
no excitations along the time axis.  He stated that this space-time
asymmetry makes it difficult to make quantum mechanics consistent with
relativity.
\par
In 1945~\cite{dir45}, Dirac attempted to use a four-dimensional
Gaussian form to construct a representation of the Lorentz group,
but did not reach specific conclusions.  In 1949~\cite{dir49}, Dirac
stated that the construction of a relativistic dynamics is equivalent
to constructing a representation of the ten-parameter Poincar\'e group.
He noted difficulties in doing so.

\par
In 1963~\cite{dir63}, Dirac published a paper on the symmetry of two
harmonic oscillators, in which he constructed a set of generators for
the deSitter group $O(3,2)$.  This paper gives an indication that it
is possible to use harmonic oscillators to construct the desired
representations for the Poincar\'e group, such as Wigner's little
groups~\cite{wig39}.  Wigner's little groups are the subgroups of
the Lorentz group which dictate the internal space-time symmetries
of particles in the Lorentz-covariant world~\cite{knp86}.
\par
In this paper, we construct standing waves which can be Lorentz-boosted
using harmonic-oscillator wave functions.  In so doing, we face the
problem of the time-separation variable.  Standing waves are of course
for bound states.

\par
For the bound-state problem, the space coordinate gives the space
separation between two constituent particles, like the Bohr radius
which measures the distance between the proton and electron in the
hydrogen atom.  Then there comes the question of the time variable in
the bound-state system. This is a time-separation variable not contained
in the Schr\"odinger picture of quantum mechanics.  On the other hand,
this time separation becomes prominent when the system is boosted, and
it becomes as big as the space separation if the system moves with a
speed close to that of light.
\par

Indeed, it is a challenge how to take into account this time separation
variable.  In order to address this question, we note first that
Dirac raised the question of time-energy uncertainty relation in 1927
when the present form of quantum mechanics was inaugurated~\cite{dir27}.
Dirac noted however that there are no excitations along the
time-like direction.
\par
In this paper, we explain in detail how this time-separation variable
can be incorporated into a Lorentz covariant formalism.  In spite of
its prominence in the covariant formalism, this variable does not
exist in the Schr\"odinger picture of quantum mechanics.  There are
no measurement theories to deal with this problem.

\par
If the variable exists and we are not able to measure it, the result
is an increase in entropy and a temperate rise.  We treat this problem
systematically with the existing rules of quantum mechanics and
relativity.  It is convenient to explain this process by using the
concepts of entanglement, decoherence, and Feynman's rest of the
universe.

\par
In Sec.~\ref{coupo}, we start with the Hamiltonian for two independent
variables.  We then couple the system by making a canonical transformation.
In Sec.~\ref{holco}, the Lorentz covariant harmonic oscillator is
presented.  This formalism takes into account Dirac's concerns mentioned
in his papers~\cite{dir27,dir45,dir49}.

\par
In Sec.~\ref{entang}, we discuss how the space-time variables become
entangled when the quantum bound state is boosted. It is shown that this
entanglement leads to an increase in entropy and a rise in temperature.
In Sec.~\ref{plazma}, it is noted that the hadron appears quite differently
when it moves rapidly than when it moves slowly.  For slow hadrons,
Gell-Mann's quark model is applicable, while Feynman's parton model is
applicable to hadrons with their speeds close to that of light.  While
observing the temperature dependence of the speed, we can explain the
quark-to-parton transition as a phase transition.

\par
In combining  quantum mechanics with special relativity, quantum field
theory occupies a very importance place.  In the Appendix, we discuss
where QFT stands with respect to our work.

%------------------------------------------------------------------------
%-------------------------------------------------------------------------

\section{Coupled Oscillators and Entangled Oscillators}\label{coupo}

Let us start with two uncoupled oscillators.  The Hamiltonian for
this system is
\begin{equation} \label{ham01}
H = \frac{1}{2}\left(p_1^2 + x_1^2\right) +
       \frac{1}{2}\left(p_2^2 + x_2^2\right),
\end{equation}
where
$$
p_1 = -i\frac{\partial}{\partial x_1}, \qquad
p_2 = -i\frac{\partial}{\partial x_2} .
$$
If we introduce the variables
\begin{eqnarray}\label{norm01}
x_{+} = \frac{x_{1} + x_{2}}{\sqrt{2}}, \qquad
  x_{-} = \frac{x_{1} - x_{2}}{\sqrt{2}}, \nonumber\\[1ex]
p_{+} = \frac{p_{1} + p_{2}}{\sqrt{2}}, \qquad
  p_{-} = \frac{p_{1} - p_{2}}{\sqrt{2}}.
\end{eqnarray}
The Hamiltonian becomes
\begin{equation} \label{ham02}
H = \frac{1}{2}\left[p_{+}^2 + x_{+}^2\right] +
\frac{1}{2}\left[p_{-}^2 + x_{-}^2\right].
\end{equation}
The resulting ground-state wave function is
\begin{equation}\label{wf00}
\psi\left(x_1,x_2\right) = \frac{1}{\sqrt{\pi}}
\exp{\left[-\frac{1}{2}\left(x_{+}^2 + x_{-}^2\right)\right]}
\end{equation}

This aspect of the two uncoupled oscillators is well known.  The
question is what happens they become coupled.

\subsection{Canonical Transformation}
\par
\begin{figure}[thb]
\centerline{\includegraphics[scale=0.5]{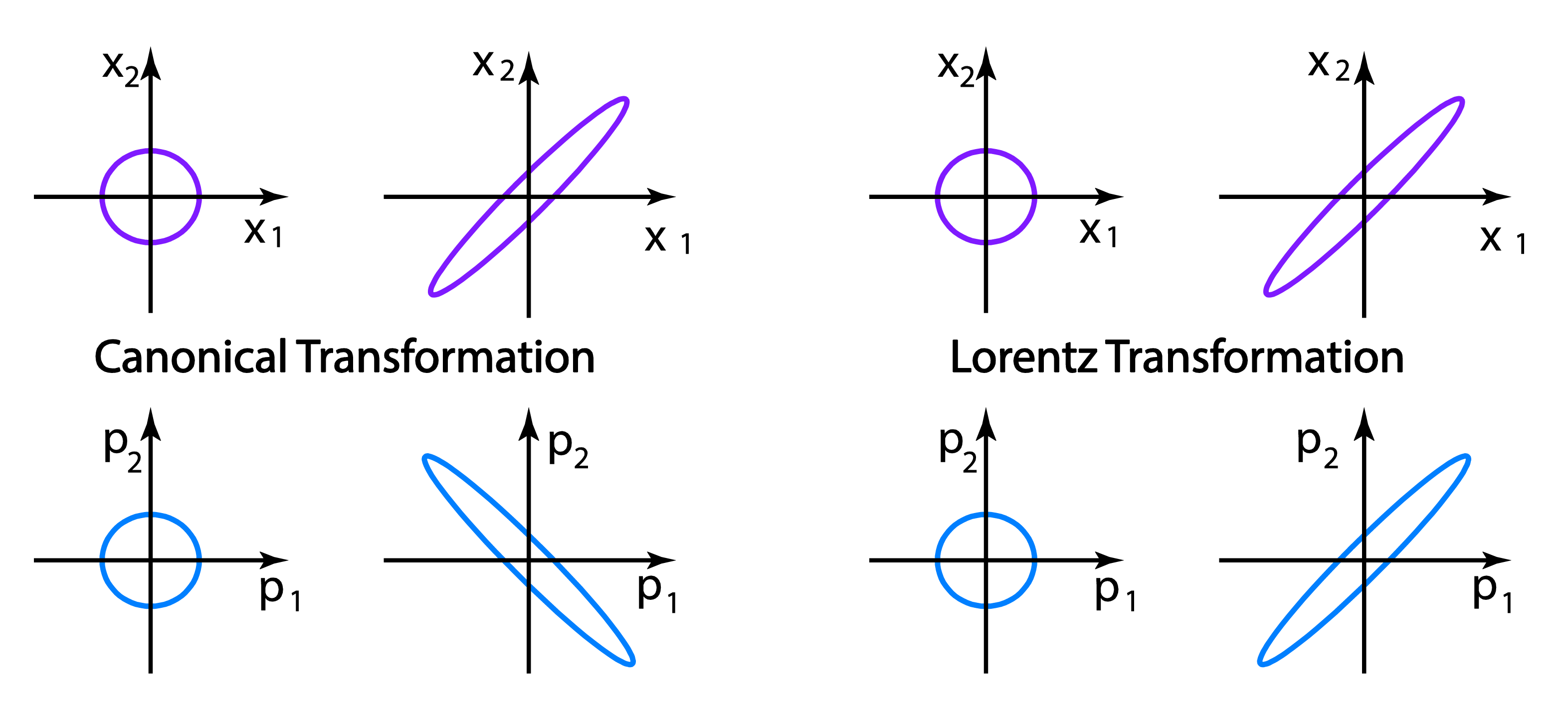}}
\vspace{5mm}
\caption{Canonical and Lorentz transformations.  In the canonical
transformation, the momentum coordinates and are inversely
proportional to their conjugate coordinates.   In the Lorentz
transformation, the momentum space and momentum coordinate transform
in the same way, as will be seen in Fig.~\ref{parton}. }\label{canlo}
\end{figure}

The canonical way to couple these two oscillators is to apply the
coordinate transformation

\begin{equation}\label{trans01}
\pmatrix{ x_{+} \cr p_{-} } \rightarrow
e^{-\eta} \pmatrix{ x_{+} \cr p_{-} } ,
\qquad
\pmatrix{x_{-} \cr p_{+} }  \rightarrow
  e^{\eta} \pmatrix{ x_{-} \cr p_{+} } .
\end{equation}
This transformation leads to the Hamiltonian of the form
\begin{equation} \label{ham03}
H_{\eta} = \frac{1}{2}\left[e^{2\eta} p_{+}^2 + e^{-2\eta}x_{+}^2\right] +
\frac{1}{2}\left[e^{-2\eta}p_{-}^2 + e^{2\eta}x_{-}^2\right],
\end{equation}
and the wave function of Eq.(\ref{wf00}) to
\begin{equation}\label{wf01}
\psi_{\eta}\left(x_1,x_2\right)
= \frac{1}{\sqrt{\pi}} \exp{\left\{-\frac{1}{4}
    \left[e^{-2\eta} \left(x_1 + x_2\right)^2
 + e^{2\eta}\left(x_1 - x_2\right)^2\right]\right\}} .
\end{equation}
This canonical transformation is illustrated in Fig.~\ref{canlo}.
From this figure, it is quite safe to say that the canonical
transformation in this case is a ``squeeze'' transformation.

\par

%------------------------------------------------------------------------
%----------------------------------------------------------------------

\subsection{Lorentz Transformation}
\par
While the Hamiltonians of Eq.(\ref{ham02}) and Eq.(\ref{ham03}) take
two different forms, we are led to the question of whether there is
a quantity invariant under this canonical transformation.  For this
purpose, let us consider the Hamiltonian of the form
\begin{equation}\label{ham05}
H_{inv} = \frac{1}{2}\left(p_1^2 + x_1^2\right) -
     \frac{1}{2}\left(p_2^2 + x_2^2\right).
\end{equation}
This is the expression for the energy of the first oscillator minus
that of the second oscillator in this two-oscillator system.  If we
write this form in terms of the $x_{\pm}$ and  $p_{\pm}$ variables,
\begin{equation}
H_{inv} = p_{+} p_{-} + x_{+} x_{-} .
\end{equation}
This form is invariant under the canonical transformation of
Eq.(\ref{trans01}).
\par
In addition, let us consider the transformation
\begin{equation}\label{trans02}
\pmatrix{ x_{+} \cr p_{+}} \rightarrow
   e^{-\eta} \pmatrix{ x_{+} \cr p_{+} } ,
\qquad
\pmatrix{ x_{-} \cr p_{-} }  \rightarrow
    e^{\eta}\pmatrix{ x_{-} \cr p_{-} } .
\end{equation}
This is not a canonical transformation.  The space and momentum
variables are transformed in the same way, as in the case of Einstein's
Lorentz boost.  Thus, we choose to call this transformation the
Lorentz transformation, and illustrated it in Fig.~\ref{canlo}.
\par
The invariant Hamiltonian was considered first by Yukawa~\cite{yuka53},
and then by Feynman {\it et al.}~\cite{fkr71}.

\subsection{Entangled Oscillators}
\par
As was discussed in the literature for several different
purposes~\cite{knp86,kno79ajp,giedke03,kn05job,kn11symm}, this wave
function can be expanded as
\begin{equation}\label{expan01}
\psi_{\eta}\left(x_{1},x_{2}\right) = {1 \over \cosh\eta}\sum^{}_{k}
(\tanh\eta )^{k} \phi_{k}(x_{1}) \phi_{k}(x_{2}) ,
\end{equation}
where $\phi_{k}(x)$ is the normalized harmonic oscillator wave function
for the $k^{th}$ excited state, and it takes the form.
\begin{equation}\label{wf02}
 \phi_{k}(x) = \left(1 \over \sqrt{\pi} n! 2^{k} \right)^{1/2}
  H_{k}(x) \exp{\left(\frac{ -x^{2}}{2}\right)} ,
\end{equation}
where $H_{k}$ is the Hermite polynomial of the $k^{th}$ order.

\par
The harmonic oscillator states are also used as photon-number states in
quantum electrodynamics.  The expansion of Eq.(\ref{expan01}) serves as
the mathematical basis for two-photon coherent states or squeezed states
of light in quantum optics~\cite{dir63,yuen76,yurke86,knp91,hkny93},
among other applications.  More recently, this expansion is called the
entangled state of two photons~\cite{giedke03}. Thus, it is appropriate
to call the expression of Eq.(\ref{expan01}) the entangled state of two
oscillators~\cite{kn05job}.

\section{Harmonic Oscillators in the Lorentz-covariant World}\label{holco}

Paul A. M. Dirac is known to us through the Dirac equation for spin-1/2
particles.  However, his main interest was in foundational problems.
First, Dirac was never satisfied with the probabilistic formulation of
quantum mechanics.  This is still one of the hotly debated subjects in
physics.  Second, if we tentatively accept the present form of quantum
mechanics, Dirac was insisting that it has to be consistent with
special relativity.  He wrote several important papers on this subject.
Let us look at some of his papers.  Among them were

\begin{itemize}

\item[1.] In 1927~\cite{dir27}, Dirac published a paper on the time-energy
  uncertainty relation.  He noted there that the time-energy relation is
  necessarily a c-number uncertainty relation because there are no
  excitations along the time-like dimension.  He noted further this
  space-time asymmetry makes it difficult to make quantum mechanics
  compatible with special relativity.

\item[2.] In his paper of 1945~\cite{dir45}, Dirac attempted to use
   harmonic oscillator wave functions to construct a representation
   of the Lorentz group.  He knew quantum mechanics is the physics of
   harmonic oscillators and special relativity is the physics of the
   Lorentz group.  For this purpose, he wrote down the Gaussian factor
\begin{equation}\label{gauss01}
\exp\left\{- {1 \over 2}\left(x^2 + y^2 + z^2 + t^2\right)\right\} .
\end{equation}
\par
Since the oscillator wave function is separable, we can consider only
\begin{equation}\label{gauss02}
\exp\left\{- {1 \over 2}\left(z^2 + t^2\right)\right\} ,
\end{equation}
when Lorentz boosts are made along the $z$ direction.  This form is
not invariant under the boost.  Furthermore, it vanishes when the $t$
variable becomes $\pm \infty.$  Does this mean that the world starts
with zero in the remote past and vanishes also in the future?

\item[3.] In his 1949 paper~\cite{dir49}, Dirac introduced the ten
  generators of the Poincar\'e group, and stated that the task of
  constructing a Lorentz-covariant quantum mechanics is equivalent to
  constructing a representation of the Poincar\'e group.  In order to
  make contact with the Schr\"odinger form of quantum mechanics, he
  considered three different forms of constraint, which are called
  instant form, front form, and point form.  Dirac then noted the
  difficulties in constructing a desired representation.
  \par
  Also in this paper, Driac introduced his light-cone coordinate
  system in order to elaborate on his front form of quantum mechanics.

\end{itemize}
\par

We propose here to construct Dirac's desired quantum mechanics by
supplementing the ideas missing in his papers.  First of all,
let us look at Gaussian form of Eq.(\ref{gauss01}).  By writing
down this Gaussian form, Dirac was clearly interested in
standing waves.  Standing waves in quantum mechanics require boundary
conditions.  The issue is then how those boundary conditions appear
to a moving observer.  We can take care of this problem if we use
harmonic oscillators with their built-in boundary conditions, as
Dirac did in his 1945 paper~\cite{dir45} without mentioning it.
\par

\begin{figure}[thb]
\centerline{\includegraphics[scale=0.7]{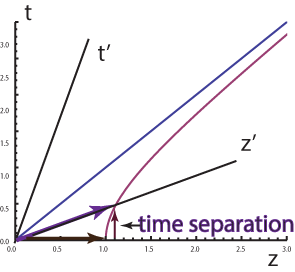}}
\vspace{1mm}
\caption{Time separation in relativity. In relativity, there is a
time separation wherever there is a space separation.  The Bohr radius
is a space separation.  It was Feynman {\it et al.}~\cite{fkr71} who
pointed out the existence of this variable, but they said they did not
know what to do with it in their oscillator formalism.}\label{tsepa}
\end{figure}

However, in writing down the Gaussian form, Dirac forgot to mention
that the space and time variables are space and time separation
variables. For instance, the Bohr radius measures the space separation
between the proton and electron.  Thus the $t$ variable in the Gaussian
form is the time separation variable which vanishes if the hydrogen
atom is at rest.  However, this time separation becomes prominent
when the hydrogen atom moves, as illustrated in Fig.~\ref{tsepa}.
\par

%------------------------------------------------------------------------
%-------------------------------------------------------------------------

Thus, the $t$ variable in the Gaussian form does not say the system
becomes zero in the remote past or remote future.  This variable
is hidden in the Schr\"odinger form of quantum mechanics.  The question
then is how to hide this variable when we extract the Schr\"odinger
picture of quantum mechanics from Dirac's Lorentz-covariant form.

\par

In his 1949 paper~\cite{dir49}, Dirac's states his constraints are
not consistent with the Poincar\'e covariance.  However, Dirac
overlooked Wigner's work on the little groups which dictate
internal space-time symmetries of particles~\cite{wig39}.
The little groups are maximal subgroups of the Lorentz group whose
transformations leave the four-momentum of a given particle invariant.
Thus, these subgroups cannot satisfy the full symmetries of the
Poincar\'e group.
\par
Let us go back to the Gaussian form of Eq.(\ref{gauss01}) again.  It
clearly indicates that there is an uncertainty relation along the time
direction, but Dirac's instant form forbids excitations along this
axis.  This aspect is perfectly consistent with the c-number time-energy
uncertainty relation Dirac mentioned in his 1927 paper~\cite{dir27}.
\par
\begin{figure}[thb]
\centerline{\includegraphics[scale=0.5]{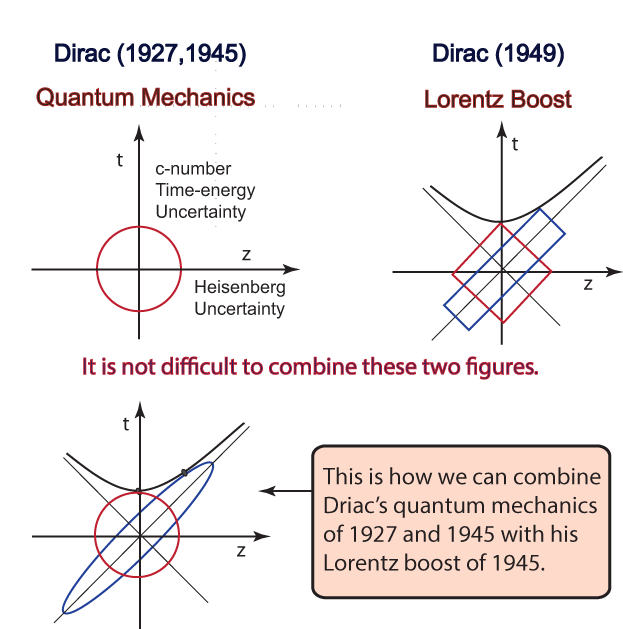}}
\vspace{2mm}
\caption{Space-time picture of bound-state quantum mechanics.  If we
combine Dirac's 1927 and 1945 papers~\cite{dir27,dir45}, we end up an
oscillator wave function without excitations along the time axis. Dirac's
1949 paper~\cite{dir49} gives a squeeze picture of the Lorentz boost. It
is then not difficult to combines these two pictures to construct a
Lorentz-covariant picture of the wave function.}\label{diracqm}
\end{figure}

If we make up for all the soft spots in Dirac's papers mentioned in this
section, we end up with an oscillator model of Lorentz-covariant
quantum mechanics illustrated in Fig.~\ref{diracqm}.

\par

%----------------------------------------------------------------------
%----------------------------------------------------------------------
\par
In Fig.~\ref{diracqm}, we start with the Gaussian expression given
in Eq.~\ref{gauss02}.  This expression is not invariant under the
Lorentz boost along the $z$ direction, which can be written as
\begin{equation}\label{boostm}
\pmatrix{z' \cr t'} =
\pmatrix{\cosh\eta & \sinh\eta \cr \sinh\eta & \cosh\eta }
\pmatrix{z \cr t } ,
\end{equation}
with
\begin{equation}\label{velo}
\tanh(\eta) = v/c ,
\end{equation}
where $v$ is the velocity of the boost.
\par
If we use Dirac's light-cone variables defined as~\cite{dir49},
\begin{equation}\label{uv01}
u = (z + t)/\sqrt{2} , \qquad v = (z - t)/\sqrt{2} ,
\end{equation}
the boost transformation of Eq.(\ref{boostm}) takes the form
\begin{equation}\label{lorensq}
u' = e^{\eta } u , \qquad v' = e^{-\eta } v .
\end{equation}
The $u$ variable becomes expanded while the $v$ variable becomes
contracted, as is illustrated in Fig.~\ref{diracqm}.  Their product
\begin{equation}
uv = {1 \over 2}(z + t)(z - t) = {1 \over 2}\left(z^2 - t^2\right)
\end{equation}
remains invariant.
\par
In terms of these variables, the Gaussian form can be written as
\begin{equation}\label{gauss03}
\exp{\left\{-{1\over 2}\left(u^{2} + v^{2}\right)\right\}} .
\end{equation}
If we apply the Lorentz boost on this function, it becomes
\begin{equation}\label{gauss05}
\exp{\left\{-{1\over 2}\left(e^{-2\eta }u^{2} + e^{2\eta}v^{2}\right)\right\}} .
\end{equation}
In this way, the harmonic oscillator wave function becomes Lorentz-boosted.
As in the case of the canonical transformation of Fig.~\ref{canlo}, the
Lorentz boost is a squeeze transformation. Thus, we can study the
Lorentz boost of the oscillator wave functions in terms of those in the
system of coupled oscillators.

\subsection{Feynman's Oscillators }\label{feynosc}
\par
\begin{figure}[thb]
\centerline{\includegraphics[scale=1.6]{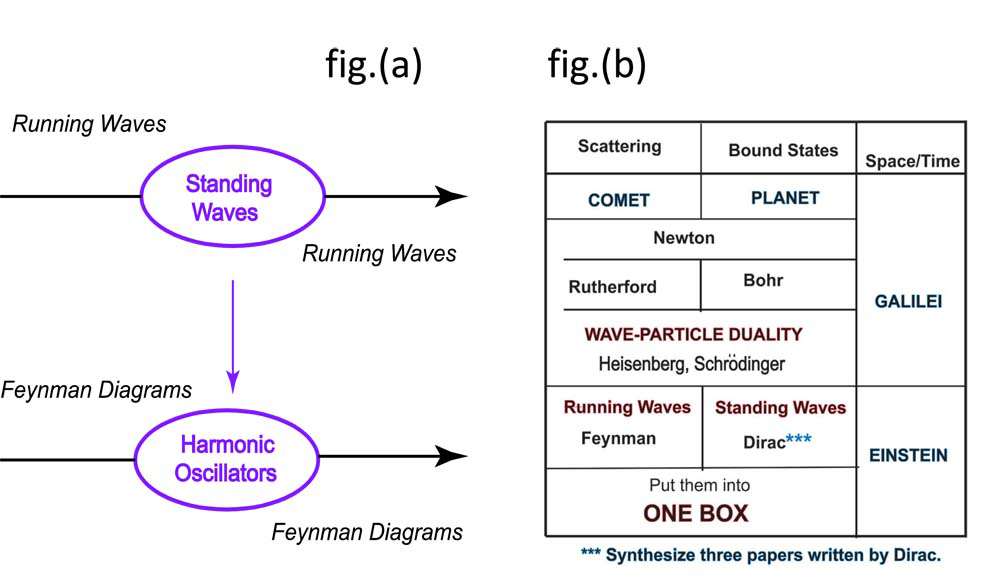}}
\vspace{1mm}
\caption{Feynman's suggestion for combining quantum mechanics with special
relativity.  Feynman diagrams work for running waves, and they provide
a satisfactory interpretation of scattering states in Einstein's
world. For standing waves trapped inside a hadron, Feynman suggested
harmonic oscillators as the first step, as illustrated in fig.(a).
From his suggestion, we can construct a historical perspective
starting from comets and planets, as shown in fig.(b). Dirac's three
papers are discussed in Sec.~\ref{holco}.  It has been shown by
Han {\it et al.} in 1981 that the covariant oscillator formalism
shares the same set of physical principles as Feynman diagrams
for scattering problems~\cite{hkn81fop}.}\label{runstan}
\end{figure}

In his invited talk at the 1970 spring meeting of the American Physical
Society held in Washington, DC (U.S.A.), Feynman noted first that the
observed hadronic mass spectra are consistent with the degeneracy of
three-dimensional harmonic oscillators.  Furthermore, he stressed
that Feynman diagrams are not necessarily suitable for relativistic bound
states and that we should try harmonic oscillators.  Feynman's point was
that, while plane-wave approximations in terms of Feynman diagrams work
well for relativistic scattering problems, they are not applicable to
bound-state problems.  We can summarize what Feynman said in
Fig.~\ref{runstan}.   The content of his talk was published in 1971
by  Feynman, Kislinger, and Ravndal~\cite{fkr71}.

\par
They use the simplest hadron consisting of two quarks bound together
with an attractive force, and consider their space-time positions $x_{a}$
and $x_{b}$, and use the variables
\begin{equation}
X = (x_{a} + x_{b})/2 , \qquad x = (x_{a} - x_{b})/2\sqrt{2} .
\end{equation}
The four-vector $X$ specifies where the hadron is located in space and
time, while the variable $x$ measures the space-time separation
between the quarks.  According to Einstein, this space-time separation
contains a time-like component which actively participates as in
Eq.(\ref{boostm}), if the hadron is boosted along the $z$ direction.
This aspect of Einstein's special relativity is illustrated in
Fig.~\ref{tsepa}.

%------------------------------------------------------------------------

%-------------------------------------------------------------------------

\par
For the internal space-time separation coordinates, Feynman {\it et al.}
start with the Lorentz-invariant differential equation~\cite{fkr71}
\begin{equation}\label{osc16}
\frac{1}{2} \left(x^{2}_{\mu} -
{\partial^{2} \over \partial x_{\mu }^{2}}
\right) \psi(x) = \lambda \psi(x) ,
\end{equation}
and write down their solutions.  We would like to point out that
those solutions are not compatible with the existing principle of
physics, but they can be modified so that they are.

\par
Since this oscillator equation is separable in the Cartesian coordinate
system, and since the transverse components are not affected by Lorentz
boost, we can drop the $x$ and $y$ components, and write Eq.(\ref{osc16})
as
\begin{equation}\label{osc17}
\frac{1}{2} \left\{\left[z^{2} - \left(\frac{\partial}{\partial z}\right)^2\right]
 - \left[t^{2} - \left(\frac{\partial}{\partial t}\right)^2\right] \right\}
 \psi(z,t) = \lambda \psi(z,t) .
\end{equation}
Here, the $z$ and $t$ variables are the space-time separation variables
respectively.  This partial differential equation has many different
solutions depending on the choice of separable variables and boundary
conditions.  Feynman {\it et al.} chose the solutions with the Gaussian
form
\begin{equation}\label{gauss07}
\exp{\left\{-{1 \over 2}\left(z^2 - t^2 \right) \right\}} ,
\end{equation}
as the ground-state wave function.  This form is invariant under
Lorentz boosts along the $z$ direction, but is not normalizable in the
$t$ variable.  For this reason, they drop this variable by setting
$t = 0$.  In so doing they are destroying the Lorentz invariance they
insisted on initially.

\par

In their paper, Feynman {\it et al.} studied in detail the degeneracy
of the three-dimensional harmonic oscillators, and compared their
results with the observed experimental data.  Their work is complete and
thorough, and is consistent with the $O(3)$-like symmetry dictated
by Wigner's little group for massive particles~\cite{wig39,knp86}.
Yet, they make an apology that the symmetry is
not that of $O(3,1)$.  This unnecessary apology causes confusion not
only to the readers but also to the authors themselves.

\subsection{Lorentz-covariant Harmonic Oscillators}\label{covham}
\par
On the other hand, as was noted earlier~\cite{yuka53},
the Gaussian form of Eq.(\ref{gauss02}) also satisfies the
Lorentz-invariant differential equation of Eq.(\ref{osc17}).
This form is not Lorentz-invariant but becomes Lorentz-squeezed
according to Eq.(\ref{gauss05}).

In order to study whether the c-number time-energy uncertainty
relation is consistent with Wigner's $O(3)$-like little group for
a massive particle, let us go back to the frame where the hadron is
at rest.  If we take into account the $x, y,$ and $z$ variables, it is
possible to construct oscillator wave functions satisfying
the $O(3)$ symmetry, with their proper angular momentum
quantum numbers.  This wave function should take the form
\begin{equation}\label{wf07}
\psi(x,y,z,t) = \psi(x,y,z)
\left[\left(\frac{1}{\pi}\right)^{1/4}
     \exp{\left(\frac{-t^{2}}{2}\right)}\right]  ,
\end{equation}
without time-like excitations~\cite{kno79jmp}.  The wave function
$\psi(x,y.z)$ can be written in terms of the spherical
variables~\cite{knp86}.  Since the oscillator equation is separable
in the Cartesian coordinate system, this wave function can also be
written in terms of the Hermite polynomials.  Here again, the
transverse variables $x$ and $y$ are not affected by the boost along
the $z$ direction, the wave function of interest can be written as
\begin{equation}\label{wf08}
 \psi^{n}(z,t) = \phi_{n}(z) \left[\left(\frac{1}{\pi}\right)^{1/4}
    \exp{\left(\frac{-t^{2}}{2}\right)}\right]   ,
\end{equation}
where $\phi_n(z)$ is for the $n^{th}$ excited oscillator state, given
in Eq.(\ref{wf02}).
When the system is at rest, the full wave function is
\begin{equation}\label{wf09}
 \psi_{0}^{n}(z,t) = \left[1 \over \pi n! 2^{n} \right]^{1/2}
  H_{n}(z)\exp{\left\{-\left(\frac{z^{2} + t^{2}}{2}\right)\right\}} .
\end{equation}
The subscript $0$ means that the wave function is for the hadron at rest.
\par
We are now interested in boosting this wave function, and it is
convenient to use the light-cone variables~given in Eq.(\ref{uv01}).
In terms of these variables, the rest-frame wave function can be
written as
\begin{equation}\label{wf10}
 \psi_{0}^{n}(z,t) = \left[1 \over \pi n! 2^{n} \right]^{1/2}
       H_{n}\left({u + v \over \sqrt{2}}\right)
       \exp{\left\{-\left(\frac{u^{2} + v^{2}}{2}\right)\right\}} ,
\end{equation}
If the system is boosted, the wave function becomes
\begin{equation}\label{wf12}
 \psi_{\eta}^{n}(z,t) = \left[1 \over \pi n! 2^{n} \right]^{1/2}
       H_{n}\left({e^{-\eta}u +  e^{\eta} v \over \sqrt{2}}\right)
   \exp{\left\{-\left(\frac{e^{-2\eta}u^{2} + e^{2\eta}v^{2}}{2}
     \right)\right\}} .
\end{equation}
\par

\begin{figure}[thb]
\centerline{\includegraphics[scale=0.6]{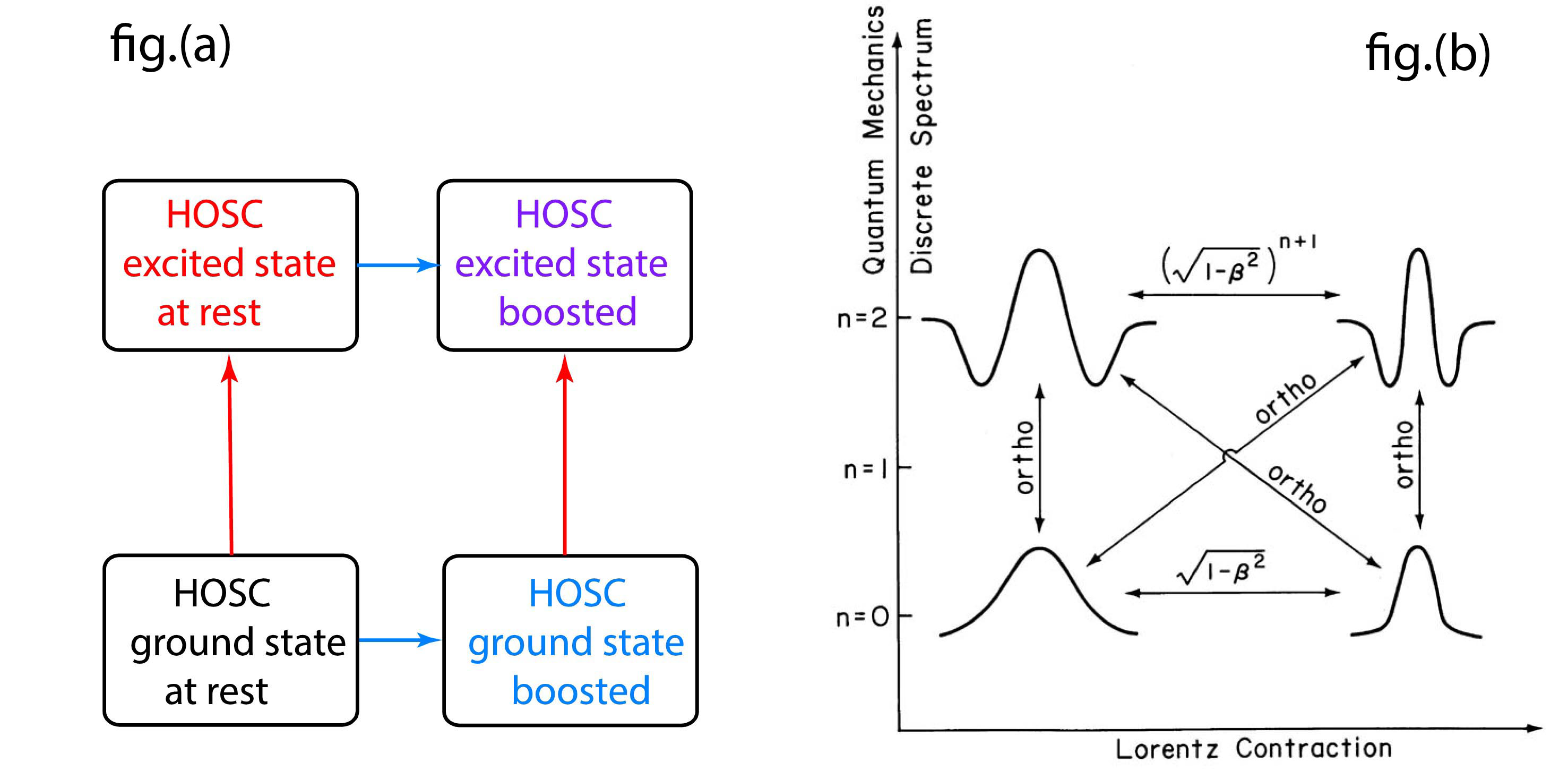}}
\caption{Orthogonality relations for covariant oscillator wave functions.
The orthogonality relations remain invariant under Lorentz boosts,
as shown in fig.(a).  Their inner products have the Lorentz-contraction
property given in fig.(b).}\label{ortho}
\end{figure}

It is interesting that these wave functions satisfy the orthogonality
condition~\cite{ruiz74}.
\begin{equation}
\int \psi^{n}_{0} (z,t) \psi^{m}_{\eta} (z,t)dz dt =
\left(\sqrt{1 - \beta^2}\right)^{n} \delta_{nm} ,
\end{equation}
where $\beta = \tanh \eta $.  This orthogonality relation is
illustrated in Fig.~\ref{ortho}.  The physical interpretation of this
in terms of Lorentz contractions is given in our book~\cite{knp86},
but it seems to require further investigation.

%----------------------------------------------------------------------
%----------------------------------------------------------------------

\section{Entangled Space and Time}\label{entang}

As was discussed in the literature for several different purposes,
the Lorentz-boosted wave function of Eq.(\ref{wf12})
can be expanded as~\cite{knp86,hkn99ajp,kiwi90pl}
\begin{equation}\label{expan02}
    \psi_{\eta}^{n}(z,t) = \left(\frac{1}{\cosh\eta}\right)^{(n+1)}
     \sum_{k} \left[\frac{(n+k)!}{n!k!}\right]^{1/2}
     (\tanh\eta)^{k}\phi_{n+k}(z)\phi_{n}(t) .
\end{equation}
For the ground state with $n = 0$, this expansion becomes
\begin{equation}\label{expan03}
\psi_{\eta }(z, t) = {1 \over \cosh\eta}\sum^{}_{k}
(\tanh\eta )^{k} \phi_{k}(z) \phi_{k}(t) ,
\end{equation}
This expansion is identical to that for the coupled oscillators if
$z$ and $t$ are replaced by $x_{1}$ and $x_{2}$ respectively.
Indeed, this is the formula for the space-time entangled state
of the ground-state oscillator with the c-number time-energy
uncertainty state.
\par
Let us go to the expansion of Eq.(\ref{expan02}).  We are
justified to call this formula the entangled state of
the $n^{th}$ excited state with the Gaussian form for
the c-number time-energy uncertainty relation.
\par
If the system is at rest, this time distribution can be
separated from the rest, and we can recover Schr\"odinger's
quantum mechanics without the time-separation variable.
If the system moves with non-zero value of $\eta$, this
time-separation variable is entangled with the Schr\"odinger
wave functions.  What should we do?

\par
In his book on statistical mechanics~\cite{fey72}, Feynman
makes the following statement about the density matrix.
\par
{\it When we
solve a quantum-mechanical problem, what we really do is divide the
universe into two parts - the system in which we are interested and
the rest of the universe.  We then usually act as if the system in
which we are interested comprised the entire universe.  To motivate
the use of density matrices, let us see what happens when we include
the part of the universe outside the system}.
\par
\begin{figure}[thb]
\centerline{\includegraphics[scale=1.4]{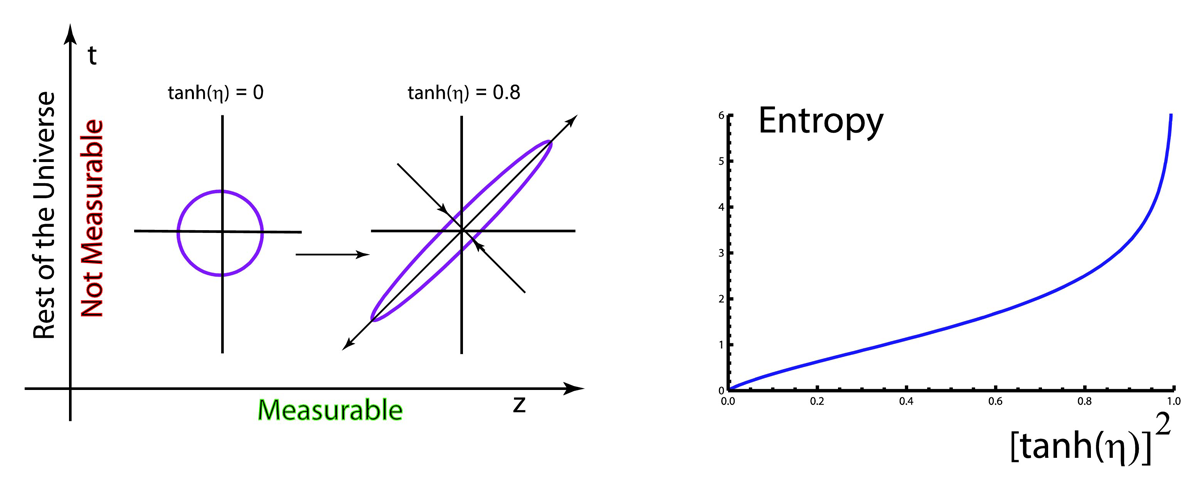}}
\vspace{2mm}
\caption{Localization property in the $zt$ plane.  When the hadron is at
rest, the Gaussian form is concentrated within a circular region specified
by $ (z + t)^2 + (z - t)^2  = 1.$  As the hadron gains speed, the region
becomes deformed to  $ e^{-2\eta}(z + t)^2  + e^{2\eta}(z - t)^2 = 1.$
Since it is not possible to make measurements along the $t$ direction,
we have to deal with information that is less than complete.  This does not
cause problems when the hadron is at rest, because the $t$ dependence in
the density matrix is separable and can be integrated out.  When the $t$
separation is not measured as in the case of the Schr\"odinger quantum
mechanics, the entropy of the system becomes non-zero and becomes
increased as the hadron gains speed.}\label{restof}
\end{figure}

Figure~\ref{restof} illustrates the time-separation variable belonging
to Feynman's rest of the universe.  If we do not observe the variable
in the rest of the universe, it causes an increase in
entropy.

%-------------------------------------------------------------------------
%-------------------------------------------------------------------------

\subsection{Entropy and Lorentz Transformations}
\par
If the $z$ and $t$ variables are both measurable, we can construct
the density matrix
\begin{equation}\label{dm01}
  \rho_\eta^{n}(z,t;z',t') =
     \psi_\eta^{n}(z,t) \left(\psi_\eta^{n} (z',t')\right)^* ,
\end{equation}
from Eq.(\ref{wf08}).  Since the wave functions in the present paper
are all real, we shall hereafter drop the $*$ sign for complex
conjugate.
\par
This form satisfies the pure-state condition
$\rho^2  = \rho$ which can be written explicitly as
\begin{equation}
  \rho_\eta^{n}(z,t;z',t') = \int \rho_\eta^{n}(z,t;z",t")
   \rho_\eta^{n}(z",t";z',t') dz"dt" .
\end{equation}

However, there are at present no measurement theories which accommodate
the time-separation variable $t$.  Thus, we can take the trace of the
$\rho$ matrix with respect to the $t$ variable.  Then the resulting
density matrix is
\begin{eqnarray}\label{dm03}
&{}& \rho_\eta^{n}(z,z') = \int \psi_\eta^{n}(z,t)
                            \psi_\eta^{n}(z',t) dt \nonumber \\[2ex]
&{}& \hspace{8mm}  = \left(\frac{1}{\cosh\eta}\right)^{2(n + 1)}
     \sum_{k} \frac{(n+k)!}{n!k!}
     (\tanh\eta)^{2k}\phi_{n+k}(z)\phi_{k+n}(z') .
\end{eqnarray}
The trace of this density matrix is one, but the trace of $\rho^2$ is
less than one, as
\begin{eqnarray}\label{dm05}
&{}& Tr\left(\rho^2\right) = \int \rho_\eta^{n}(z,z')
                        \rho_\eta^{n}(z',z) dz dz' \nonumber \\[2ex]
  &{}& \hspace{10mm} = \left(\frac{1}{\cosh\eta}\right)^{4(n + 1)}
  \sum_{k} \left[\frac{(n+k)!}{n!k!}\right]^2 (\tanh\eta)^{4k} .
\end{eqnarray}
which is less than one.  This is due to the fact that we do not
know how to deal with the time-like separation in the present
formulation of quantum mechanics.  Our knowledge is less than
complete.
\par
The standard way to measure this ignorance is to calculate the
entropy defined as~\cite{neu32,fano57,landau58,wiya63}
\begin{equation}
          S = - Tr\left(\rho \ln(\rho)\right)  .                            %(16)
\end{equation}
If we pretend to know the distribution along the time-like direction and
use the pure-state density matrix given in Eq.(\ref{dm01}), then the entropy is
zero.  However, if we do not know how to deal with the distribution along
t, then we should use the density matrix of Eq.(\ref{dm03}) to calculate the
entropy, and the result is~\cite{kiwi90pl}
\begin{eqnarray}
&{}&  S = (n + 1)\left[\left(\cosh^2\eta\right) \ln\left(\cosh^2\eta\right) -
              (\sinh^2\eta)\ln\left(\sinh^2\eta\right)\right] \nonumber \\[1ex]
&{}& \hspace{10mm} - \left(\frac{1}{\cosh\eta}\right)^{2(n + 1)}
  \sum_{k} \frac{(n+k)!}{n!k!}\ln\left[\frac{(n+k)!}{n!k!}\right]
               (\tanh\eta)^{2k} .
\end{eqnarray}
In terms of the velocity $v$ of the hadron,
\begin{eqnarray}
&{}&  S = -(n + 1)\left\{\ln\left[1 - \left(\frac{v}{c}\right)^2\right]
      + \frac{(v/c)^2 \ln(v/c)^2}{1 - (v/c)^2}   \right\}    \nonumber \\[2ex]
&{}& \hspace{10mm} - \left[1 - \left(\frac{1}{v}\right)^2\right]
  \sum_{k} \frac{(n+k)!}{n!k!}\ln\left[\frac{(n+k)!}{n!k!}\right]
               \left(\frac{v}{c}\right)^{2k} .
\end{eqnarray}

\par
For the ground state with $n = 0$, the density matrix of Eq.(\ref{dm03})
becomes
\begin{equation}\label{dm07}
\rho_{\eta}(z,z') = \left(\frac{1}{\cosh\eta}\right)^{2}
     \sum_{k} (\tanh\eta)^{2k}\phi_{k}(z)\phi^*_{k}(z') ,
\end{equation}
and the entropy becomes
\begin{equation}
  S = \left(\cosh^2\eta\right)\ln\left(\cosh^2\eta\right) -
        \left(\sinh^2\eta\right) \ln\left(\sinh^2\eta\right)  .
\end{equation}

\par

Let us go back to the wave function given in Eq.(\ref{wf12}).  As is
illustrated in Fig.~\ref{restof}, its localization property is
dictated by the Gaussian factor which corresponds to the ground-state
wave function.  For
this reason, we expect that much of the behavior of the density matrix or
the entropy for the n-th excited state will be the same as that for the
ground state with n = 0.  For the ground-state, the wave function
becomes
\begin{equation}\label{wf18}
 \psi_{\eta}(z,t) = \left[1 \over \pi \right]^{1/2}
   \exp{\left\{-\left(\frac{e^{-2\eta}u^{2} + e^{2\eta}v^{2}}{2}
     \right)\right\}} .
\end{equation}
\par
For this state, the density matrix is
\begin{equation}
  \rho(z,z') = \left(\frac{1}{\pi \cosh(2\eta)}\right)^{1/2}
  \exp{\left\{-\frac{1}{4}\left[\frac{(z + z')^2}{\cosh(2\eta)}
                    + (z - z')^2\cosh(2\eta)\right]\right\}} ,
\end{equation}
The quark distribution $\rho(z,z)$ becomes
\begin{equation}\label{eq21}
  \rho(z,z) = \left(\frac{1}{\pi \cosh(2\eta)}\right)^{1/2}
  \exp{\left(\frac{-z^2}{\cosh(2\eta)}\right) }.
\end{equation}
The width of the distribution becomes $\sqrt{\cosh(2\eta)}$, and
becomes wide-spread as the hadronic speed increases.  Likewise, the
momentum distribution becomes wide-spread~\cite{knp86,kimli89pl}. This
simultaneous increase in the momentum and position distribution
widths is called the parton phenomenon in high-energy physics~\cite{fey69a,fey69b}.
We shall return to this problem in Sec.~\ref{plazma}.

\subsection{Hadronic Temperature}
\par
Harmonic oscillator wave functions are used for all branches of physics.
The ground-state harmonic oscillator can be excited in the following three
different ways.
\begin{itemize}
\item[1.] Energy level excitations, with the energy eigenvalues
          $\hbar \omega (n + 1/2)$.

\item[2.] Coherent state excitations resulting in
           \begin{equation}
  |\alpha> = e^{\alpha a^{\dagger}}
      = \sum_{n} \frac{\alpha^n}{\sqrt{n!}}|n> .
           \end{equation}
\item[3.] Thermal excitations resulting in the density matrix of the
          form
\begin{equation}\label{dm09}
  \rho_{T}(z, z') = \left(1 - e^{-\hbar\omega/kT}\right)
  \sum_{k} e^{-k\hbar\omega/kT} \phi_{k}(z)\phi_{k}^{*}(z'),
    \end{equation}
where $\hbar\omega$ and $k$ are the oscillator energy separation
and Boltzmann's constant respectively. This form of the density
matrix is well known~\cite{landau58,kimli89pl,davies75,hkn90pl}.

\end{itemize}

\par
\begin{figure}[thb]
\centerline{\includegraphics[scale=0.6]{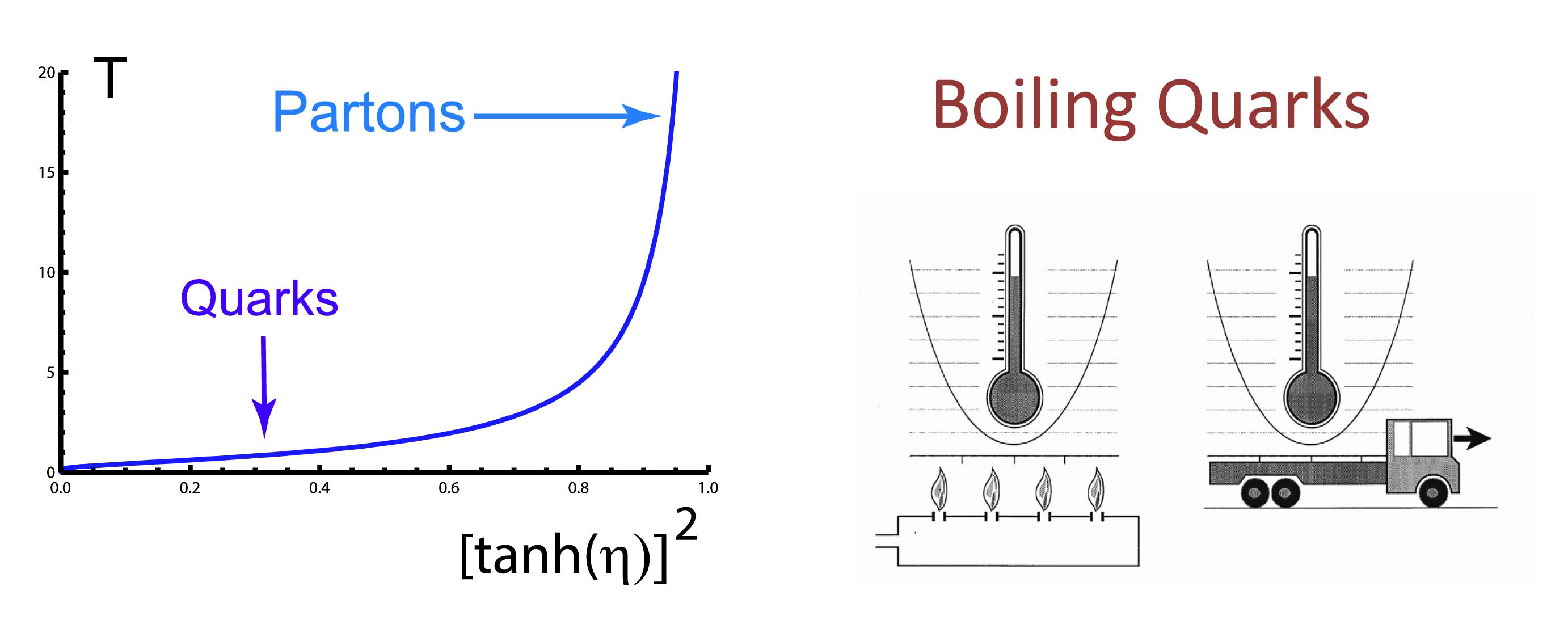}}
\vspace{1mm}
\caption{Boiling quarks become partons.  When the hadron gains
speed, the temperature of system rises according to Eq.(\ref{temp02}).
The quarks will boil, and they will go through a phase transition
to partons as is indicated in Fig.~\ref{phasetr}.}\label{qboil}
\end{figure}

In this subsection, we are interested in the thermal excitation.
If the temperature is measured in units of $\hbar\omega/k$,
the density matrix of Eq.(\ref{dm09}) can be written as
\begin{equation}\label{densi16}
  \rho_{T}(z, z') = \left(1 - e^{-1/T}\right)
  \sum_{k} e^{-1/T} \phi_{k}(z)\phi_{k}^{*}(z'),
    \end{equation}
If we compare this expression with the density matrix of Eq.(\ref{dm03}),
we are led to
\begin{equation}\label{temp01}
\tanh^2\eta  = \exp{\left(-1/T \right)},
\end{equation}
and to
\begin{equation}\label{temp02}
T = \frac{-1}{\ln{\left(\tanh^2\eta\right)}}
\end{equation}
The temperature can be calculated as a function of $\tanh(\eta)$,
and this calculation is plotted in Fig.~\ref{qboil}.

\par
Earlier in Eq.(\ref{velo}), we noted that $\tanh(\eta)$ is
proportional to velocity of the hadron, and $\tanh(\eta) = v/c$.
Thus, the oscillator becomes thermally excited as it moves, as is
illustrated in Fig.~\ref{qboil}.

%---------------------------------------------------------------------------
%---------------------------------------------------------------------------

\par
Let us look at the velocity dependence of the temperature
again.  It is almost proportional to the velocity from
$\tanh(\eta) = 0 $ to $0.7$, and again from $\tanh(\eta) = 0.9$
to $1$ with different slopes.  We shall return to this issue
in Sec.~\ref{plazma} in connection with the transition from the
quark model to the parton model.
\par
While the physical motivation for this section was based on Feynman's
time separation variable~\cite{fkr71} and his rest of the
universe~\cite{fey72}, we should note that many authors discussed
field theoretic approach to derive the density matrix of
Eq.(\ref{dm09}).  Among them are two-mode squeezed states of
light~\cite{yuen76,yurke86,knp91,hkny93} and
thermo-field-dynamics~\cite{fewa71,ume82,mann89}.
\par
The mathematics of two-mode squeezed states is the same as that for
the covariant harmonic oscillator formalism discussed in this
paper~\cite{dir63,yurke86,knp91,hkny93}.  Instead of the $z$ and
$t$ coordinates, there are two measurable photons.  If we choose
not to observe one of them, it belongs to Feynman's rest of the
universe~\cite{hkn99ajp,yupo87,ekn89}.

\par

Another remarkable feature of two-mode squeezed states of light is
that its formalism is identical to that of
thermo-field-dynamics~\cite{fewa71,ojima81,ume82,mann89}.
The temperature is in this case related to the squeeze parameter.  It
is therefore possible to define the temperature of a Lorentz-squeezed
hadron within the framework of the covariant harmonic oscillator model.

\section{Quark-Parton Phase Transitions}\label{plazma}

\begin{figure}[thb]
\centerline{\includegraphics[scale=1.8]{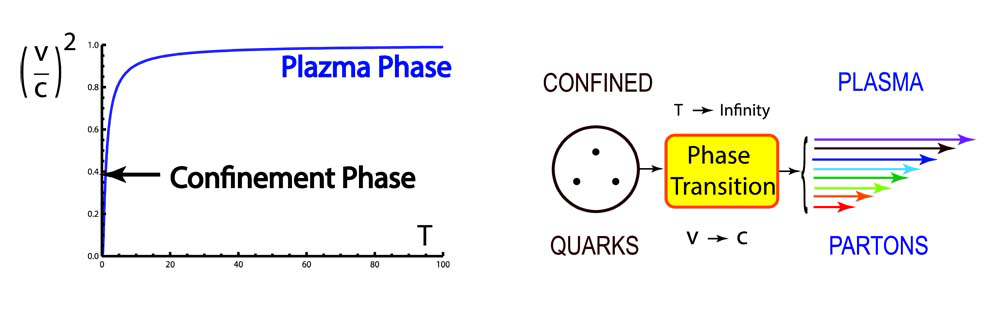}}
\caption{Transition from the confinement phase to a plasma phase.
The quarks are confined within a hadron when the hadron is at
rest, but they behave like free particles when the hadron speed
reaches that of light, and the temperature becomes very high.  This
figure is a extended interpretation of Fig.~\ref{qboil}.}\label{phasetr}
\end{figure}

Let us go back to Fig.(\ref{qboil}).  There the hadronic temperature $T$
is plotted against $\tanh^2(\eta) $ or $(v/c)^2$.  We can also plot
$(v/c)^2$ as a function of $T$, as shown in Fig.~\ref{phasetr}.

\par

As is seen in Fig.~\ref{phasetr}, the curve is nearly vertical
for low temperature, but becomes nearly horizontal for
high temperature, even though it is continuous.  Thus, we are
led to suspect a phase transition between these two different
sections of the curve.  Let us look at what happened inside the
hadron.
\par
If the hadron is at rest or its speed is very low, we use the
quark model.  If the hadron is moving with the speed close to
that of light, we use the parton model.  Since the constituents
behave quite differently in these two models, we are confronted
with the question of whether they can be described as two different
limiting cases of one covariant entity.
\par
This kind of question is not new in physics.  Before 1905, Einstein
faced the question of two different energy-momentum relations for
massive and massless particles.  He ended up with the formula
$E = \sqrt{m^2 + p^2} $, which is widely known as his $E = mc^2$.
\par
Our quark-parton puzzle is similar to that of energy-momentum
relation, as illustrated in Table~\ref{emc2}.  The dynamics of
the quark model is the same as that of the hydrogen atom, where
two constituent particles are bound together by an attractive force.
\par
We are all familiar with the dynamics of the hydrogen atom, but
it is a challenge to understand Feynman's parton picture as a
Lorentz-boosted hydrogen atom.  The key variable is the
time-separation between the quarks, not seen in the Schr\"odinger
picture of the hydrogen atom.  This variable has been studied
in detail in Sec.~\ref{holco}.

\par

%---------------------------------------------------------------------------
%---------------------------------------------------------------------------
\par

%--------------------------------------------------------------------------
\begin{table}%[thb]
\caption{Massive and massless particles in one package.  Einstein
unified the energy-momentum relation for slow (massive) and fast
(massless) particles with one Lorentz-covariant formula.}\label{emc2}
\vspace{1mm}
\begin{center}
\begin{tabular}{lcccccc}
\hline\\[2mm]
{} &{}& Massive &{}& Lorentz &{}& Massless \\
{} &{}& Slow  &{}& Covariance &{}& Fast \\[2mm]
\hline
{}&{}&{}&{}&{}&{}\\
Energy-  &{}&  $E =$      &{}& Einstein's &{}  \\
Momentum &{}&  $p^{2}/2m$ &{}& $ E = [p^{2} + m^{2}]^{1/2}$ &{}& $E = p$
   \\[4mm]
\hline
{}&{}&{}&{}&{}&{}\\
Hadron's &{}& Gell-Mann's  &{}& Lorentz-covariant &{}& Feynman's\\
Constituents &{}& Quark Model & {}& Harmonic Oscillator  &{}& Parton Picture
    \\[4mm]\hline
\end{tabular}
\end{center}
\end{table}
%------------------------------------------------------------------------

\subsection{Feynman's Parton Picture}\label{fparton}
\par
In a hydrogen atom or a hadron consisting of two quarks, there is a
spacial separation between two constituent elements.  In the case of
the hydrogen atom we call it the Bohr radius.  If the atom or hadron is
at rest, the time-separation variable does not play any visible role
in quantum mechanics.  However, if the system is boosted to the
Lorentz frame which moves with a speed close to that of light, this
time-separation variable becomes as important as the space separation
of the Bohr radius.  Thus, the time-separation plays a visible role
in high-energy physics which studies fast-moving bound states.  Let
us study this problem in more detail.
\par
It is a widely accepted view that hadrons are quantum bound states
of quarks having a localized probability distribution.  As in all
bound-state cases, this localization condition is responsible for
the existence of discrete mass spectra.  The most convincing evidence
for this bound-state picture is the hadronic mass
spectra~\cite{fkr71,knp86}.
However, this picture of bound states is applicable only to observers
in the Lorentz frame in which the hadron is at rest.  How would the
hadrons appear to observers in other Lorentz frames?
\par
In 1969, Feynman observed that a fast-moving hadron can be regarded
as a collection of many ``partons'' whose properties appear
to be quite different from those of the quarks~\cite{fey69a,fey69b}.  For
example, the number of quarks inside a static proton is three, while
the number of partons in a rapidly moving proton appears to be infinite.
The question then is how the proton looking like a bound state of
quarks to one observer can appear different to an observer in a
different Lorentz frame?  Feynman made the following systematic
observations.

\begin{itemize}

\item[a.]  The picture is valid only for hadrons moving with
   velocity close to that of light.

\item[b.]  The interaction time between the quarks becomes dilated,
    and partons behave as free independent particles.

\item[c.]  The momentum distribution of partons becomes widespread as
    the hadron moves fast.

\item[d.]  The number of partons seems to be infinite or much larger
     than that of quarks.

\end{itemize}

\noindent Because the hadron is believed to be a bound state of two
or three quarks, each of the above phenomena appears as a paradox,
particularly b) and c) together.  How can a free particle have a
wide-spread momentum distribution?

In order to resolve this paradox, let us construct the
momentum-energy wave function corresponding to the Gaussian form of
Eq.(\ref{gauss05}).
If the quarks have the four-momenta $p_{a}$ and $p_{b}$, we can
construct two independent four-momentum variables~\cite{fkr71}
\begin{equation}
P = p_{a} + p_{b} , \qquad p = \sqrt{2}(p_{a} - p_{b}) .
\end{equation}
The four-momentum $P$ is the total four-momentum and is thus the
hadronic four-momentum.  $p$ measures the four-momentum separation
between the quarks.  Their light-cone variables are
\begin{equation}\label{conju}
p_{u} = \frac{p_{0} - p_{z}}{\sqrt{2}} ,  \qquad
p_{v} = \frac{p_{0} + p_{z}}{\sqrt{2}} ,
\end{equation}
with the Lorentz-boost property
\begin{equation}
p_{u} \rightarrow e^{-\eta}p_{u} \qquad
p_{v} \rightarrow e^{\eta}p_{v} .
\end{equation}
\par
The momentum-energy wave function is
\begin{equation}\label{wf20}
\phi_{\eta}(p_{z},p_{0}) = \frac{1}{2\pi}\int \psi_{\eta}(z,t)
       \exp{\left\{i\left(z p_{z} - t p_{0}\right)\right\}}dz dt
\end{equation}
where $\psi_{\eta}$ is the space-time wave function of Eq.(\ref{wf18}).
The evaluation of this integral leads to
\begin{equation}\label{phi}
\phi_{\eta }(p_{z},p_{0}) = \left({1 \over \pi }\right)^{1/2}
\exp\left\{-{1\over 2}\left[e^{-2\eta}p_{u}^{2} +
e^{2\eta}p_{v}^{2}\right]\right\} .
\end{equation}
Because we are using here the harmonic oscillator, the mathematical
form of the above momentum-energy wave function is the same as that
of the space-time wave function of Eq.(\ref{wf12}) in its ground
state.

\par

\begin{figure}[thb]
\centerline{\includegraphics[scale=0.6]{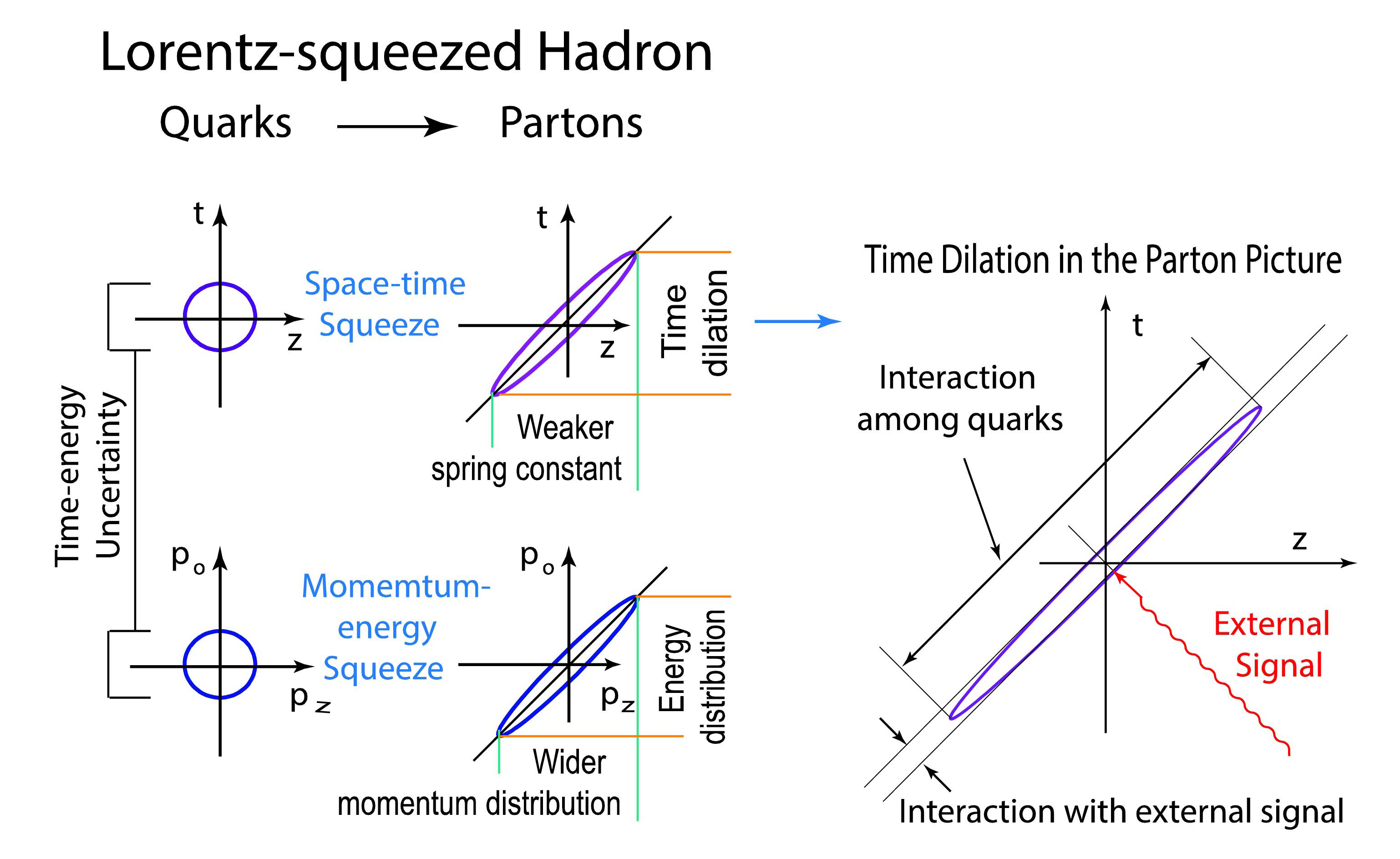}}
\vspace{1mm}
\caption{Lorentz-squeezed space-time and momentum-energy wave functions.
As the hadron's speed approaches that of light, both wave functions
become concentrated along their respective positive light-cone axes.
These light-cone concentrations lead to Feynman's parton picture.  The
interaction time of the quarks among themselves becomes dilated, as the
expanding major axis of space-time ellipse indicates.  On the other hand,
the external signal, since it is moving in the direction opposite to the
direction of the hadron, travels along the negative light-cone axis.
To the external signal, if it moves with velocity of light, the hadron
appears very thin, and the quark's interaction time with the external
signal becomes very small.  This is called Feynman's time
dilation}\label{parton}
\end{figure}

The Lorentz squeeze properties of these wave functions are also the
same, as is indicated in Fig.~\ref{canlo}.  This aspect of the
squeeze has been exhaustively discussed in the
literature~\cite{knp86,kn77par,kim89}, and they are illustrated again
in Fig.~\ref{parton} of the present paper.  The hadronic structure
function calculated from this formalism is in a reasonable agreement
with the experimental data~\cite{hussar81}.

%----------------------------------------------------------------------
%----------------------------------------------------------------------

\par
When the hadron is at rest with $\eta = 0$, both wave functions behave
like those for the static bound state of quarks.  As $\eta$ increases,
the wave functions become continuously squeezed until they become
concentrated along their respective positive light-cone axes.  Let us
look at the z-axis projection of the space-time wave function.  Indeed,
the width of the quark distribution increases as the hadronic speed
approaches that of the speed of light.  The position of each quark
appears widespread to the observer in the laboratory frame, and the
quarks appear like free particles.
\par
The momentum-energy wave function is just like the space-time wave
function.  The longitudinal momentum distribution becomes wide-spread
as the hadronic speed approaches the velocity of light.  This is in
contradiction with our expectation from nonrelativistic quantum
mechanics that the width of the momentum distribution is inversely
proportional to that of the position wave function.  Our expectation
is that if the quarks are free, they must have a sharply defined
momenta, not a wide-spread distribution.
\par
However, according to our Lorentz-squeezed space-time and momentum-energy
wave functions, the space-time width and the momentum-energy width increase
in the same direction as the hadron is boosted.  This is of course an
effect of Lorentz covariance.  This is indeed the resolution of one of the
quark-parton puzzles~\cite{knp86,kn77par,kim89}.

\subsection{Feynman's Decoherence}\label{fdeco}
\par
Another puzzling problem in the parton picture is that partons appear
as incoherent particles, while quarks are coherent when the hadron
is at rest.  Does this mean that the coherence is destroyed by the
Lorentz boost?  The answer is known to be NO~\cite{kn05job,kn03os},
and we would like to expand our earlier discussion on this subject.
\par
When the hadron is boosted, the hadronic matter becomes squeezed and
becomes concentrated in the elliptic region along the positive
light-cone axis.  The length of the major axis becomes expanded by
$e^{\eta}$, and the minor axis is contracted by $e^{-\eta}$.
\par

This means that the interaction time of the quarks among themselves
become dilated.  Because the wave function becomes wide-spread, the
distance between one end of the harmonic oscillator well and the other
end increases.  This effect, first noted by Feynman~\cite{fey69a,fey69b},
is universally observed in high-energy hadronic experiments.  The
period of oscillation increases like $e^{\eta}$.

\par
On the other hand, the external signal, since it is moving in the
direction opposite to the direction of the hadron travels along
the negative light-cone axis, as illustrated in Fig.~\ref{parton}.

\par
If the hadron contracts along the negative light-cone axis, the
interaction time decreases by $e^{-\eta}$.  The ratio of the interaction
time to the oscillator period becomes $e^{-2\eta}$.  The energy of each
proton coming out of CERN's LHC is $4000 GeV$.  This leads to a ratio of
$1.6\times 10^{-8}$.  This is indeed a small number.  The external signal
is not able to sense the interaction of the quarks among themselves inside
the hadron.

\par
Indeed, Feynman's parton picture is one concrete physical example
where the decoherence effect is observed.  As for the entropy, the
time-separation variable belongs to the rest of the universe.  Because
we are not able to observe this variable, the entropy increases as the
hadron is boosted to exhibit the parton effect.  The decoherence is
thus accompanied by an entropy increase.

\par
Let us go back to the coupled-oscillator system.  The light-cone
variables in Eq.(\ref{uv01}) correspond to the normal coordinates in
the coupled-oscillator system given in Eq.(\ref{norm01}).  According
to Feynman's parton picture, the decoherence mechanism is determined
by the ratio of widths of the wave function along the two normal
coordinates.

\par
This decoherence mechanism observed in Feynman's parton picture is quite
different from other decoherences discussed in the literature.  It is
widely understood that the word decoherence is the loss of coherence
within a system.  On the other hand, Feynman's decoherence discussed in
this section comes from the way the external signal interacts with the
internal constituents.

\subsection{Uncertainty Relations}
\par
Let us go back to Fig.~\ref{parton}.  Both the spatial and momentum
distribution become widespread as the hadron is Lorentz-boosted.
Does this mean that Plank's constant is not Lorentz invariant?  The
answer is No.  The uncertainty relation still remains invariant.
According to Eq.(\ref{phi}), the major axis of the space-time
ellipse is conjugate to the minor axis of the momentum-ellipse.
Thus the uncertainty products
\begin{equation}
\left<u^2\right>\left<p_{u}^2\right> =
 \left<v^2\right>\left<p_{v}^2\right>
\end{equation}
remain invariant under the boost.
\par
\begin{figure}[thb]
\centerline{\includegraphics[scale=1.5]{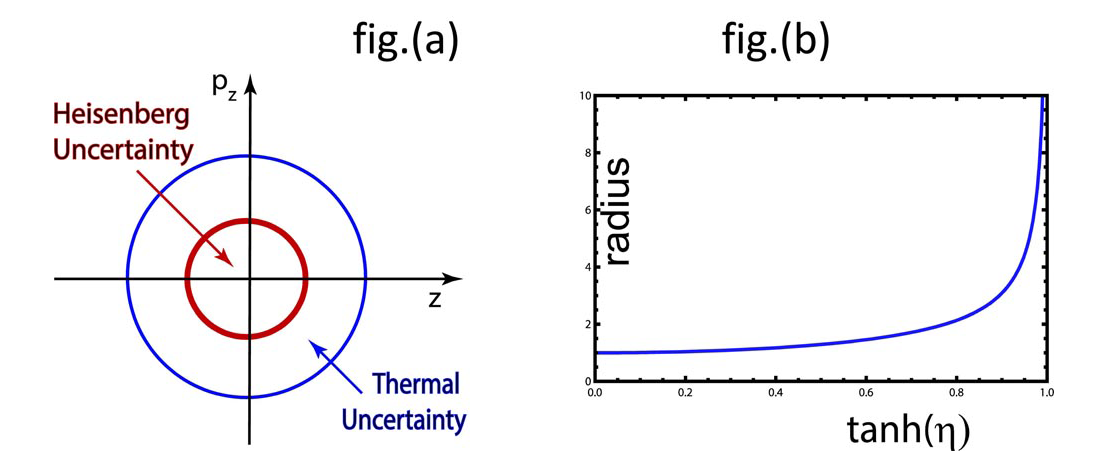}}
\vspace{5mm}
\caption{Uncertainty Distribution in the Wigner phase space.  The
radius takes the minimal value determined by the uncertainty principle
when the hadron is at rest and its temperature is zero.  The radius
increases as the temperature rises as indicated in fig.(a).  The
dependence of this radius on $\tanh(\eta)$ (the hadronic speed) is
also given in fig.(b). }\label{wigradius}
\end{figure}

What happens when we fail to observe the time-separation coordinate?
We can approach this problem using the Wigner function~\cite{wig32}.
For the density matrix of Eq.(\ref{dm01}), the Wigner function is
\begin{equation}
W_{\eta} \left(z,t,p_{z},p_{0}\right) = \frac{1}{\pi}
\int \exp{\left[(i(z'p_{z} + t' p_{0})\right]}
    \rho_{\eta}(z - z', t - t') dz' dt'
\end{equation}
The evaluation of this integral is straight-forward~\cite{knp91,davies75}.
For $\eta = 0$,
the Wigner function becomes
\begin{equation}\label{wigf01}
W_{0}\left(z,p_{z}; t,p_{0}\right) = \left(\frac{1}{\pi}\right)^{2}
   \exp{\left[-\left(z^2 + p_{z}^2 + t^2 + p_{0}^2\right)\right]} .
\end{equation}
This is the Wigner function for the minimal uncertainty state without
any thermal effects, or for the hadron at rest.  This Gaussian form
is illustrated in Fig.~\ref{wigradius}.

%----------------------------------------------------------------------
%-----------------------------------------------------------------------

\par
For non-zero values of $\eta, W_{\eta}\left(z,p_{z}; t,p_{0}\right)$
becomes
\begin{equation}
\left(\frac{1}{\pi}\right)^{2}
\exp\left\{-{1\over 2}\left[e^{2\eta} (t + z)^{2} + e^{-2\eta}(t - z)^{2}
 + e^{-2\eta}\left(p_{z}-p_{0}\right)^{2} +
 e^{2\eta}\left(p_z + p_{0}\right)^{2} \right] \right\}.
\end{equation}
\par
If we do not observe the second pair of variables, we have to
integrate this function over $t$ and $p_0$, and the resulting
Wigner function is
\begin{equation}
W_{\eta}\left(z, p_z\right) = \int W(z,p_{z};t,p_{0}) dt dp_{0} ,
\end{equation}
and the evaluation of this integration leads to~\cite{hkn99ajp}
\begin{equation}
W_{\eta}(x, p) =  \frac{1}{\pi \cosh\eta}
\exp{\left[- \left(\frac{z^2 + p_{z}^2}{\cosh(2\eta)}\right)\right] },
\end{equation}
\par
The failure to make measurements on the time separation variable
leads to a radial expansion of the Wigner phase space as in the
case of the thermal excitation.  The radius is
\begin{equation}
\sqrt{\cosh(2\eta)} = \sqrt{\frac{1 + \tanh^2\eta}
                                 {1 - \tanh^2\eta}} .
\end{equation}
As is indicated in Fig.~\ref{wigradius}, the radius becomes larger
when $\tanh(\eta)$ becomes larger or the hadron moves with an
increasing speed.

\section*{Concluding Remarks}

For forty years since 1973~\cite{kn73}, the present authors published
many papers on the subject of Lorentz-boosting standing waves.  What
is new in this paper?

\par
In our first paper on this subject~\cite{kn73}, we started with a
Lorentz-covariant Gaussian form proposed by Yukawa's in his 1953
paper~\cite{yuka53}, and its applications to high-energy
physics~\cite{markov56,ginz65,fuji70,licht70,lipes72}.  We were
particularly interested in how the Gaussian form becomes deformed
under Lorentz boosts.
\par
Since then we have been adding physical interpretations to Yukawa's
Gaussian form.  In 1977~\cite{kn77par}, using the Yukawa form, we
were able to show that the quark model and the parton model are two
different manifestations of the same covariant model.
\par
In 1979, using a set of normalizable solutions of the Lorentz-invariant
differential equation of Feynman~\cite{fkr71}, which includes Yukawa's
Gaussian form, we constructed representations of Wigner's $O(3)$-like
little group for massive particles~\cite{kno79jmp}.  Wigner's little
groups dictate the internal space-time symmetries of particles in the
Lorentz-covariant world.

\par
During the period of 1980-81, Kim and Han noted that the wave function
approach starting from Yukawa is equivalent to Dirac's plan to construct
a Lorentz-covariant quantum mechanics~\cite{hk80ptp,hk81ajp}.  According
to these papers, the covariant harmonic oscillator wave functions
combine Dirac's efforts made in his three papers~\cite{dir27,dir45,dir49}
into Fig.~\ref{diracqm}.

\par
In so doing we had to address the physics of the time separation between
the two constituent particles.  This variable does not exist in the
Schr\"odinger picture of quantum mechanics, while the time-energy
uncertainty relation plays the key role in calculation of transition
rates~\cite{dir27}.  The covariant harmonic oscillator embraces these
two physical principles.
\par
However, the failure to observe this time-separation variable leads
to a rise in entropy and temperature.  It is shown that this can be
interpreted in terms of the space-time entanglement.  In this paper,
it was noted that the Lorentz boost entangles the longitudinal and
time-like variables.  We have addressed the issue of the entropy and
temperature arising from this entanglement.

\par
In Feynman's parton picture, interaction of the quarks with the external
signal should be coherent according to quantum mechanics.  However, they
lose coherence when they become partons.  This is another physical
example of decoherence.
\par

The mathematics of the covariant harmonic oscillators is basically the
same as that of the coupled harmonic oscillators.  Its mathematics is
transparent, and its quantum mechanics is well understood.  It is
gratifying to note that this simple mathematical instrument is capable
of embracing many of the physical concepts developed in recent years,
such as squeezed states, entanglement, decoherence, as well as the
following issues raised by Feynman.

\begin{itemize}

\item[1.]  Three-dimensional harmonic oscillators are quite adequate
           for describing the observed hadron spectra~\cite{fkr71}.

\item[2.]  We have to use harmonic oscillators, instead of Feynman
           diagrams, for bound-state problems in the Lorentz-covariant
           world.  In so doing we have to deal with the problem of the
           time-separation variable, while Feynman did not~\cite{fkr71}.

\item[3.]  In 1969~\cite{fey69a,fey69b}, Feynman proposed his parton
           model for hadrons  with their speeds close to the speed of
           light.  Since the partons appear to be quite different from  <---
           the quarks inside the hadrons at rest, the question is <---
           whether Feynman's parton model is a Lorentz-boosted quark
           model, or the quark and parton models are two different
           manifestations of one Lorentz-covariant entity. We addressed
           this issue in Subsec.~\ref{fparton}.

\item[4.]  Feynman's rest of the universe~\cite{fey72}.  The
           time-separation variable exists according to Einstein.
           Since we are not making observations on this variable,
           it belongs to Feynman's rest of the universe.  We
           discussed in detail what happens if we do not observe
           this variable.

\end{itemize}

In this paper, we have addressed these Feynman issues by combining
the three papers of Dirac mentioned in Sec.~\ref{holco}.

Finally, on the subject of combining quantum mechanics with special
relativity, quantum field theory occupies an prominent place.  The
question then is what role QFT plays in the problems discussed in this
paper.  We address this issue in the Appendix.

\begin{appendix}
\section*{Appendix}
In this appendix, we would like discuss why we did not use the present
form of quantum field theory, while it combines quantum mechanics
with special relativity.  Yes, quantum electrodynamics produces the Lamb
shift and the correction to the electron magnetic moment.   However, in
order to produce the Lamb shift, we need the Rydberg formula for the hydrogen
energy levels.  Does the present form of QFT produce a Lorentz-covariant
picture of hydrogen energy levels?  The answer clearly NO.  QFT does not
solve all the problems.

We are not the  first ones to see this difficulty.  While Feynman was the
person who formulated Feynman diagrams for QFT, he suggested the use of
harmonic oscillators to approach tackle bound-state problems in his talk
at the 1970 Washington meeting of the American physical society~\cite{fkr71}.
Since 1973~\cite{kn73}, we rigorously followed Feynman's suggestion
and have published a series of papers.

\par
Feynman's suggestion was based on the Chew-Frautschi plot of the hadronic
mass spectra~\cite{chew62}.  It was noted that they are reflections of
degeneracies of the three-dimensional harmonic oscillators.
Feynman {\it et al.} then wrote down a Lorentz-invariant partial
differential equation for the harmonic oscillator.  However, they could
not produce the solution consistent with Lorentz covariance and
the probability interpretation quantum mechanics.

\par
The most serious error in this paper is their treatment of the time
separation variable between the quarks.  Indeed, our present paper
is dedicated to a better understanding of this unobservable variable
in terms of Feynman's rest of the universe~\cite{fey72}.
\par
In a series of publications, we have constructed solutions of their
oscillator equation consistent with Wigner's little groups dictating
internal space-time symmetries, resulting in our 1986 book~\cite{knp86},
and consistent with Dirac's c-number time-energy uncertainty
relation~\cite{dir27}.
\par
The most significant result of our endeavor is to show that Gell-Mann's
quark model and Feynman's parton model are two limiting case of one
Lorentz-covariant entity.  The purpose of the present paper is to
elaborate on this point.
\par
The question still is where our program stands with respect to quantum
field theory.  Let us go to Fig.~\ref{runstan}.  There we point out there
that there are running and standing waves in quantum mechanics corresponding
to scattering and bound states.  They may have different forms of
mathematics, but we point out they should and they do share the same
set of physical principles of quantum mechanics and special
relativity~\cite{hkn81fop}.

\end{appendix}

\bibliographystyle{unsrt}

\end{document}